\documentclass{article}[margin=1cm]
\usepackage{graphicx} % Required for inserting images
\usepackage{amsmath}
\usepackage{amssymb}
\usepackage[margin=2.5cm]{geometry}
\usepackage{titling}
\usepackage{float}
\usepackage{caption}
\usepackage{subcaption}
\usepackage{booktabs}
\usepackage{amsthm}
\usepackage{color,soul}
\usepackage{tikz}
\usepackage{titlesec}
\usepackage{appendix}
\usepackage{titlesec}
\titleclass{\subsubsubsection}{straight}[
  \subsubsection]
\newcounter{subsubsubsection}[subsubsection]
\renewcommand\thesubsubsubsection{\thesubsubsection.
oman{subsubsubsection}}
\titleformat{\subsubsubsection}[runin]
  {\normalfont\normalsize\bfseries}{\thesubsubsubsection}{1em}{}
\titlespacing*{\subsubsubsection}{0pt}{3.25ex plus 1ex minus .2ex}{1em}

\newtheorem{example}{Example}

\title{Learning the Word Problem: Geodesic Lengths and Cryptographic Applications.}
\author{Elisabeth Fink \\
\small Middlesex University, Department of Computer Science, The Burroughs, Hendon, London NW4 4BT, UK \\
\small \texttt{e.fink@mdx.ac.uk}}
\date{\today}

\begin{document}

\maketitle

\begin{abstract}
The Word Problem has been a subject of intensive mathematical study for over a century, initially driving advances in combinatorial group theory and more recently emerging as a foundational hardness assumption in post-quantum cryptography (PQC). While generally undecidable, several families of infinite non-abelian groups exhibit solvable or algorithmically fast word problems, making them attractive platforms for cryptographic design. This paper introduces \emph{WPNet}, a novel Graph Neural Network architecture capable of solving the Word Problem heuristically, which is demonstrated on the Baumslag-Solitar group $BS(1,2)$ and on an Artin group. By mapping unreduced words to dynamic graph structures, the model learns to cluster algebraically equivalent elements in a continuous embedding space, effectively identifying the geodesic representative of a word without executing discrete reduction steps. As an application, a model variant is developed that can predict the geodesic length of an unreduced word in both groups. To demonstrate the cryptographic severity of this structural leakage, WPNet is successfully deployed against the Wagner-Magyarik public-key cryptosystem. 
\end{abstract}

\section{Introduction}

The word problem, introduced by Dehn in 1911, is the foundational decision problem in combinatorial group theory: given a finite presentation of a group and a word in its generators, determine whether the word represents the identity element \cite{Dehn1911}. It is undecidable in general \cite{Novikov1955, Boone1958}. A natural refinement is the geodesic (minimal reduced-word) length problem -- computing the shortest word representing a given element with respect to a fixed generating set -- which underlies growth series and the metric geometry of Cayley graphs. While automatic groups admit regular languages of geodesics and rational geodesic growth series, no efficient exact algorithm for geodesic length is known for many groups of interest, including braid groups, other Artin groups, and Baumslag--Solitar groups such as BS(1,2): computing the geodesic length in a subgroup is formally hard \cite{MyasnikovUshakov2008}, and the set of geodesics in the infinite braid group is co-NP-complete.

This intractability is not merely of theoretical interest: it is the mathematical basis for post-quantum group-based cryptography. With Shor's algorithm threatening standard public-key systems based on factorization and discrete logarithms \cite{Shor1997}, non-abelian protocols have emerged as a quantum-resistant alternative. The Wagner-Magyarik cryptosystem, one of the earliest such schemes, encodes messages by masking bits within tangled identity words, relying directly on the word problem's hardness \cite{wagner1984public}. Despite being broken by chosen-ciphertext and reaction attacks \cite{levy2005wagner, vasco2004wagner}, its direct-decryption vulnerability makes it a natural benchmark for evaluating neural word-problem solvers. The Anshel-Anshel-Goldfeld (AAG) and Ko-Lee protocols instead rely on the Conjugacy Search Problem in braid groups \cite{Anshel1999, Ko2000}, requiring legitimate parties to solve the word problem to verify key alignment \cite{Myasnikov2008}. Crucially, protocol security depends on choosing obscuring word lengths that resist length-based attacks: efficient approximate length oracles have been shown to untangle conjugated words of several hundred generators in braid groups \cite{Garber2006, Myasnikov2006}, forcing modern designs toward word lengths in the thousands, where the exponential (or worse) Dehn function of the underlying group is relied upon to make direct search intractable \cite{Gersten1992, Platonov2004}. Any advance in efficiently solving the word or geodesic length problem -- algorithmic or learned -- therefore directly bears on which groups remain viable post-quantum cryptographic platforms, and at what word length would be required.

This tension between algorithmic tractability and geometric complexity is exemplified by two benchmark families: The Baumslag-Solitar group $BS(1,2)$ and Artin groups with $m_{i,j}=2,3$.  Baumslag-Solitar groups $BS(m,n) = \langle a, t \mid t^{-1}a^m t = a^n \rangle$ were introduced as the first examples of two-generator, one-relator non-Hopfian groups \cite{Baumslag1962}. For the solvable case $BS(1,q)$, the word problem lies in $\mathrm{LOGSPACE}$ and uniform $TC^0$ \cite{Weiss2016, Ganardi2020}, yet the Dehn function is strictly exponential for $|q| \ge 2$ \cite{Gersten1992} -- a dichotomy between cheap algorithmic decidability and exponential geometric area that makes $BS(1,2)$ an especially demanding benchmark for a model that must recover geodesic representatives rather than merely decide triviality. Elements in $BS(1,2)$ have a normal form and the length of a geodesic representing such an element can be computed in linear time (\cite{Elder2010Geodesics}). Artin groups present a complementary challenge: the word problem is solved for several subclasses via tailored geometric and rewriting techniques -- spherical type via normal forms \cite{Artin1947, GARSIDE1969, Brieskorn1972, Deligne1972, Adian1984}, right-angled and FC type via geometric methods \cite{Charney1992biautomatic, VanWyk1994, Hermiller1995, Altobelli1998}, large and extra-large type via small cancellation and shortlex automaticity \cite{Appel1983, Peifer1996, HoltRees2012}, and affine, locally reducible, and 3-free types via related rewriting arguments \cite{Digne2006, McCammond2017, Dehornoy2013}, but no solution is known in general, and groups mixing braid and commutator relations of the kind studied here fall outside all of these solved subclasses. The two types of groups were chosen to have $BS(1,2)$ as one representative that is computationally very well understood and with the Artin group $A(\Gamma)$ another group, where the word problem is not known to be decidable.

Machine learning has recently been applied to problems of this kind. Models have been trained to predict permutations and infer relations directly from data in symmetric groups \cite{petschack2025learning}, and architectures such as AlgebraNets embed associative-algebra structure natively into network parameters \cite{hoffmann2020algebra}. Closest to the present setting, decision trees, random forests, and neural classifiers have solved the conjugacy decision problem in polycyclic and metabelian groups, including BS(1,2) itself \cite{GryakHaralickKahrobaei2020}. A separate line of work learns distance directly on a group's Cayley graph: a deep network trained to estimate cost-to-go on the Rubik's cube Cayley graph via approximate value iteration guides A* search to near-optimal solutions \cite{Agostinelli2019}, an approach since reformulated as graph neural network node classification \cite{Barro2025} and scaled to arbitrary finite permutation groups via diffusion-distance estimation \cite{Chervov2025RL, Chervov2025Cube}. Equivariant Fourier-basis heuristics offer a further design axis \cite{PanKondor2021} (see \cite{JuDong2026} for a broader survey). Reinforcement learning has separately been used to untangle braids via Reidemeister moves \cite{Lisitsa2022}. However, all of this prior work operates at word lengths far below the thousands of generators required for cryptographic relevance: graph neural networks trained on shorter words face a fundamental scaling barrier, as models operating on word topologies exceeding their receptive field fail to propagate algebraic information globally, while increasing depth to accommodate $L \ge 1000$ induces severe over-smoothing \cite{Li2018, Oono2020, Alon2021}. No prior work has targeted geodesic length prediction for unreduced words in Artin groups with mixed braid and commutator relations at cryptographically relevant lengths, nor evaluated learned heuristics against the specific hardness landscape of $BS(1,2)$ and braid-type presentations.

This paper investigates the applicability of graph neural networks to the word problem at scale. In particular, variations of WPNet are developed to evaluate triplets of words in Artin and BS groups: each word -- reduced or unreduced -- is embedded in the 128-dimensional unit sphere $\mathcal{S}^{128}$, and given a triplet $(w, p, n)$ where $w$ is an unreduced word, $p$ a geodesic representative of $w$, and $n$ a word derived from $p$ by small perturbations, WPNet is deemed to have learned a geodesic representation of $w$ if the embeddings of $w$ and $p$ lie close together while those of $w$ and $n$ are well separated. The present results show successful separation of true geodesic representatives from decoys with over 90\% accuracy for word lengths up to 5000 in both Artin and BS groups. Further, models to predict geodesic length in $BS(1,2)$ and this Artin group are developed, achieving 99\% accuracy up to word length 200. Finally, this approach is applied to the Wagner-Magyarik system, recovering the correctly passed message with near-100\% accuracy even when hidden in words up to length 1000.

\section{Methods}

\subsection{Groups and preliminaries}

Let $G$ be a finitely presented group defined by $G = \langle S \mid R \rangle$, where $S$ is a finite set of generators and $R$ is a set of defining relations. Elements of $G$ are represented by words formed from the generators and their formal inverses, $S \cup S^{-1}$. The relations are words in $S \cup S^{-1}$ and determine cancellations, in other words, which words equate to the identity. The length of a word $w$ is the number of letters required to represent the element. In this work, the length of a word will frequently refer to the length before reduction using group relations.

The word problem asks whether a given word $w$ represents the identity element in $G$. The complexity of reducing a trivial word to the identity is captured by the Dehn function, $\delta(n)$. It measures the maximum number of times the relations in $R$ must be applied to mathematically reduce any trivial word of length at most $n$ back to the empty word. 

A group $G$ is metabelian if its commutator subgroup $H = \langle g^{-1}h^{-1}gh | g,h \in G\rangle$ is abelian. An HNN extension (\cite{higman1949embedding}) constructs a larger group from a base group $H$ by introducing a stable letter $t$ that enforces an isomorphism between two subgroups of $H$. A group $G$ is defined as Hopfian (\cite{hopf1931beitrage}) if every surjective endomorphism $f: G \to G$ is an isomorphism. Equivalently, $G$ is Hopfian if it is not isomorphic to any of its proper quotient groups, that is, $G \not\cong G/N$ for any non-trivial normal subgroup $N \triangleleft G$.

Let $G = \langle S \mid R \rangle$ be a finitely presented group. For any word $w$ over $S \cup S^{-1}$ that represents the identity element in $G$, the area of $w$, denoted $\operatorname{Area}(w)$, is the minimal integer $k$ such that $w$ can be expressed in the free group as a product of $k$ conjugates of the defining relations and their inverses: 
$$ w = \prod_{i=1}^k u_i r_i^{\pm 1} u_i^{-1} $$
The Dehn function $\delta: \mathbb{N} \to \mathbb{N}$ of the given finite presentation is defined as the maximum area required to reduce any trivial word of length at most $n$:
$$ \delta(n) = \max \left\{ \operatorname{Area}(w) \mid w =_G 1 \text{ and } |w| \le n \right\} $$
It acts as an isoperimetric inequality, bounding the worst-case geometric complexity of the word problem for the group (\cite{gromov1987hyperbolic}).

\subsubsection{The Baumslag-Solitar Group $BS(1,2)$}

The first target algebraic structure in this study is the Baumslag-Solitar group $BS(1,2)$, due to its well-understood word problem paired with exponential growth. This group is  defined by the finite presentation $$BS(1,2) = \langle a, b \mid b a b^{-1} a^{-2} = 1 \rangle.$$ The broader family of Baumslag-Solitar groups, denoted $BS(m,n) = \langle a, b \mid b a^m b^{-1} = a^n \rangle$, was originally introduced by Gilbert Baumslag and Donald Solitar in 1962 to provide the first examples of two-generator, one-relator non-Hopfian groups (specifically when $m$ and $n$ share no common prime divisors, such as in $BS(2,3)$) \cite{Baumslag1962}. 

While $BS(1,2)$ is Hopfian, it has become a central benchmark in combinatorial and geometric group theory due to the severe dichotomy between its algorithmic space complexity and its geometric area complexity. The major mathematical results defining the behavior of $BS(1,2)$ are broadly categorized into algorithmic solvability and isoperimetric growth. 

Because $BS(1,2)$ is an HNN extension of the integers and belongs to the class of metabelian groups, its Word Problem is globally decidable. Algorithmically, it can be solved with extremely high efficiency. Recent complexity results have demonstrated that the Word Problem for $BS(1, q)$ can be solved within the highly restrictive uniform $\text{TC}^0$ complexity class, which defines a highly restrictive model of parallel computation (\cite{vollmer1999introduction}). Being in $\text{TC}^0$ imposes strict upper bounds on both the sequential and parallel time complexity of the word problem in $BS(1,2)$. Sequentially, evaluating a uniform $\text{TC}^0$ circuit requires time proportional to its size, meaning the word problem is solvable in deterministic polynomial time ($\mathcal{O}(n^c)$ for some constant $c$) \cite{Weiss2016, ganardi2020knapsack, Ganardi2021}.

In contrast to its algorithmic efficiency, the geometric landscape of $BS(1,2)$ is notoriously hostile. The Dehn function, $\text{Area}(W)$, measures the worst-case number of times the defining relator ($bab^{-1}a^{-2}=1$) must be applied to mathematically reduce a trivial word $W$ of length $N$ back to the identity element. Gersten proved that the Dehn function for $BS(1,2)$ exhibits strictly exponential growth \cite{Gersten1992}.

This exponential scaling is precisely what makes $BS(1,2)$ an optimal test environment for evaluating Graph Neural Networks in cryptographic contexts. A naive, greedy string-rewriting algorithm, or an attacker attempting a brute-force length-based search, will be quickly overwhelmed by the exponentially exploding number of intermediate reduction steps required to untangle a long identity word. 

\subsubsection{Stochastic Artin Groups with Mixed Commutation and Braid Relations}

To evaluate the capacity of the architecture, the model is tested against a stochastically generated Artin group. While the Baumslag-Solitar group tests the network's ability to navigate exponential geometric areas, this Artin group evaluates the model's capacity to learn high-dimensional rewriting systems derived from arbitrary Coxeter graphs. In Artin groups relations take the form:\begin{equation}\underbrace{x_i x_j x_i \dots}_{m_{i,j} \text{ terms}} = \underbrace{x_j x_i x_j \dots}_{m_{i,j} \text{ terms}}\end{equation} The Coxeter graph $G = (V, E)$ provides a compact encoding of the Artin group’s algebraic structure, where each vertex $v \in V$ corresponds to a generator $x_i$. Edges between vertices are labeled by integers $m_{i,j} \ge 2$. The word problem for Artin groups is only partially understood, but remains open in general, in particular for Artin groups with 3-relations ($m_{i,j} = 3$).

The construction of random groups by defining relations probabilistically relies on the Erdős-Rényi random graph model $\Gamma(n, p)$. The group is defined over a large rank of generators, $S = \{x_0, x_1, x_2, \dots, x_{100}\}$, with the group presentation governed by a random symmetric adjacency matrix $M$. Such constructions were defined and used in \cite{artin1925theorie, baudisch1977subgruppen, brieskorn1971fundamentalgruppe, tits1961groupes}.

The experimental setup generates a sparse interaction matrix for the Coxeter graph where some of the generator pairs ($10\%$) do not interact ($m_{i,j} = \infty$), hence generate a free subgroup. The remaining pairs ($90\%$) are distributed between $m_{i,j}=2$ and $m_{i,j} = 3$. At first, the adjacency matrix of the Coxeter graph is filled with $0,2,3$, such that each entry has a 10\% probability to be 0 (representing $\infty$), a 45\% probability to be $2$ or $3$ respectively. Coxeter graphs have to be undirected, hence the obtained (unsymmetric) adjacency matrix is symmetrized by setting $a_{i,j} = max(a_{i,j}, a_{j,i})$, hence prefering braid relations over commutators. With a chosen random seed this led to 44 pairs of generators having no relation, 1523 pairs of generators commuting and 3483 pairs of generators having a braid relation.

Such frameworks have been widely utilized to determine the topological properties, hyperbolicity thresholds, and automorphism group finiteness of random right-angled Artin groups ($m_{i,j} \in \{2,\infty\}$) and Coxeter groups (\cite{Costa2011, Charney2010, Day2011, Behrstock2017}). Recent theoretical extensions by Goldsborough and Vaskou \cite{Goldsborough2023} have generalized this to random Artin groups by simultaneously scaling the group rank and the specific probability distributions of the permitted edge coefficients.  Right-angled Artin groups (RAAGs) are widely studied, see for example \cite{charney2007introduction, abrams2013pushing, wade2015lower, aramayona2016first, dani2018large,day2019relative, day2021calculating, kropholler2022incoherence}). For RAAGs, the Word Problem is highly tractable. Hermiller and Meier demonstrated that RAAGs possess finite, complete string-rewriting systems, allowing the Word Problem to be solved in linear time with respect to the word length \cite{Hermiller1995}.

The inclusion of braid relations ($m_{i,j} = 3$) significantly complicates the geometry. While the Word Problem for general Artin groups remains famously open, subclasses characterized by sparse interaction graphs, such as large, extra-large, and FC-type Artin groups, have been proven decidable. Holt and Rees established practical, polynomial-time solutions for wide classes of these groups using shortlex automatic structures and localized geometric reductions \cite{Holt2012}.

Due to their robust algorithmic solvability but highly complex, non-abelian subgroups, Artin groups (particularly Braid groups and RAAGs) have been heavily proposed as platforms for post-quantum key exchange protocols, making their geometric topologies prime targets for cryptanalysis \cite{Myasnikov2008, habeeb2012secret, kahrobaei2015algorithmic, flores2019cryptography}.

In this experimental setup, the WPNet must implicitly learn the underlying Coxeter graph routing. It must recognize when local sub-words can commute past each other or braid through one another to reach the same global topological state.

\subsection{Training data}\label{sec:sub_graph}

Each word in these generators is transformed by mapping discrete sequences of group generators into structured graphs. This transformation is achieved through dynamic graph builders that translate unreduced words into specialized graph structures tailored to the underlying group geometry. 

Given an unreduced word $W = (g_1, g_2, \dots, g_L)$ where $g_k \in \pm S$, first the word is broken into subwords of length 5, where a new subword starts at every even index. If the final subwords have length less than 5, they are padded with 0, which represents the identity in the respective group. Each such subword represents one node in the graph. Each letter $g$ is mapped into $\mathbb{R}^7$:
\[g \mapsto \left(\frac{g}{|S|}, \sin(g), \cos(g), \sin\left(\frac{g}{10}\right), \cos\left(\frac{g}{10}\right), sign(g), \frac{|g|}{|S|}\right).\]

Each node in the graph has a feature tensor $t \in \mathbb{R}^{5x7}$, for the $5$ letters in the subword, which is projected into a smaller space by the BiLSTM defined in the next section. A word of length $n$ will result in a graph of $\left\lceil \frac{n}{2} \right\rceil$ nodes.

\begin{figure}[H]
    \centering
    \includegraphics[width=1\linewidth]{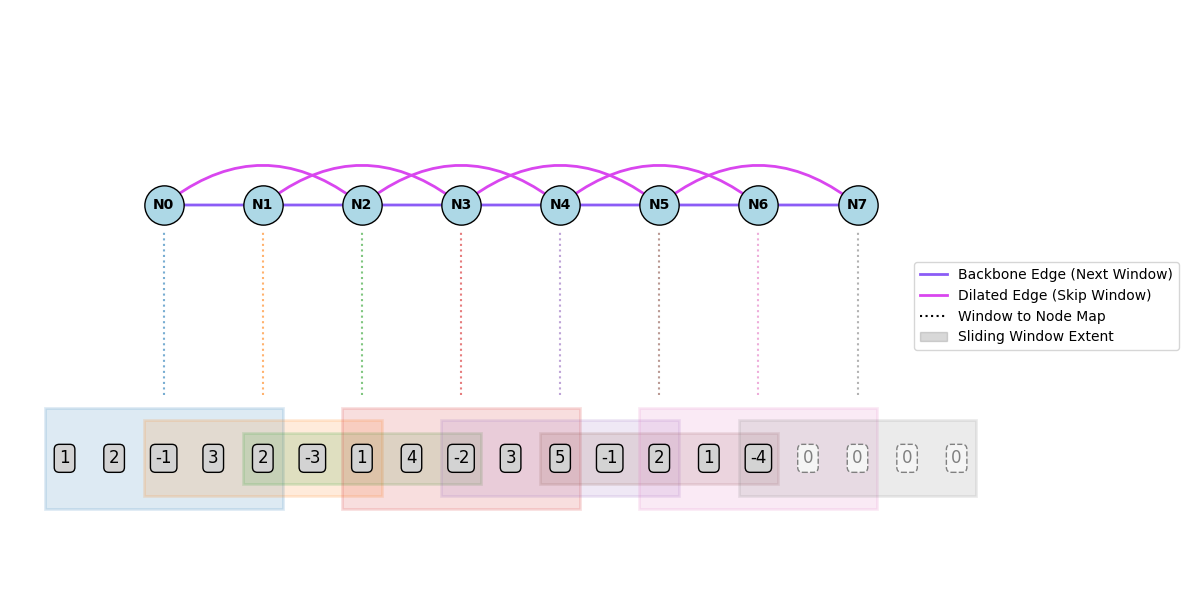}
    \caption{Graph representation of the word $w=g_1 g_2 g_1^{-1} g_3 g_2 g_3^{-1} g_1 g_4 g_2^{-1} g_3 g_5 g_1^{-1} g_2 g_1 g_4^{-1}$ in $A(\Gamma)$.}
    \label{fig:graph_representation}
\end{figure}

The resulting nodes are then connected by edges in the following way: Nodes representing neighboring subwords are connected by an edge of type 1, and nodes of distance 2 in this resulting graph are then further connected by an edge of type 2. The resulting graph for the word $w=g_1 g_2 g_1^{-1} g_3 g_2 g_3^{-1} g_1 g_4 g_2^{-1} g_3 g_5 g_1^{-1} g_2 g_1 g_4^{-1}$ in $A(\Gamma)$ is shown in Figure \ref{fig:graph_representation}, where each integer $i$ represents $g_i$.

The objective of WPNet is to determine algebraic equivalence between word representations within $BS(1,2)$ or a randomly generated Artin group of rank $101$. The training is restricted to sequences with initial lengths ranging from $L=20$ to $100$. Within the free monoid over $200$ symbols (generators and their formal inverses), the number of base configurations exceeds $200^{20} \approx 10^{46}$. This leads to a total state space of unique graph topologies that is computationally practically infinite, rendering a static approach with a fixed training table infeasible. Although a smaller, discrete subset could be pre-computed, utilizing a fixed dataset would likely induce overfitting, where the neural network memorizes specific syntactic patterns rather than true algebraic invariants. By employing a statistical learning strategy \cite{schroff2015facenet, you2020graph}, the model is exposed to a continuous variety of isometrically equivalent presentations. This forces the architecture to learn the intrinsic geometric structure of the group element rather than static sequence data. To achieve this, the training dataset consisting of 8192 sequence triplets is dynamically regenerated every five epochs to effectively prevent overfitting. 

To force the neural network to learn underlying algebraic invariants rather than relying on superficial length heuristics, the training data engine employs a stochastically driven triplet generator. The batch distribution is deliberately engineered to expose the model to the exact types of heavy cryptographic padding utilized in Wagner-Magyarik attacks. 

\subsubsection{Contrastive Learning for Word Embeddings}

The new batch generated every epoch comprises 8192 triplets $(A, P, N)$, evenly partitioned into trivial identity bases and uniformly sampled random sequence bases. The Anchor ($A$) and Positive ($P$) are constructed by independently tangling the base word through $N_{ops}$ randomized algebraic operations (e.g., inserting trivial pairs or applying group relators) to preserve mathematical equivalence. The tangling parameter $N_{ops}$ is assigned probabilistically: identity bases receive moderate to massive tangling (up to 150 operations), while random bases are tangled proportionally to their length, with a subset undergoing extreme tangling. To construct the Negative ($N$), the base word is first rendered algebraically inequivalent via internal block permutations or boundary truncations. Crucially, this mutated base is then obfuscated using the exact same $N_{ops}$ parameter applied to $A$ and $P$. This symmetrical tangling ensures $N$ perfectly matches the visual complexity and length distribution of the anchor, preventing the network from exploiting superficial length heuristics and forcing it to map deep algebraic invariants.

\begin{example}

To illustrate the generation pipeline, consider a short random base element over the generating set $\{a, b\}$.

\textbf{1. Base Words Initialization}
\begin{itemize}
    \item \textbf{Base:} $W_{base} = aba^{-1}$
    \item \textbf{Negative Base:} The base word undergoes a boundary truncation mutation (dropping the suffix generator $a^{-1}$), yielding the inequivalent word $W_{mut} = ab$.
\end{itemize}

\textbf{2. Symmetrical Tangling}
Exactly two randomized algebraic operations (in this case, the insertion of trivial freely-reducing pairs $gg^{-1}$) are independently applied to expand each word.

\begin{itemize}
    \item \textbf{Anchor ($A$):} Tangling $W_{base}$
    \begin{enumerate}
        \item Insert $bb^{-1}$ at index 1: $a \mathbf{bb^{-1}} ba^{-1}$
        \item Insert $a^{-1}a$ at index 0: $\mathbf{a^{-1}a} a b b^{-1} b a^{-1}$
    \end{enumerate}
    
    \item \textbf{Positive ($P$):} Tangling $W_{base}$ independently
    \begin{enumerate}
        \item Insert $aa^{-1}$ at index 3: $aba^{-1} \mathbf{aa^{-1}}$
        \item Insert $b^{-1}b$ at index 2: $ab\mathbf{b^{-1}b}a^{-1}aa^{-1}$
    \end{enumerate}
    
    \item \textbf{Negative ($N$):} Tangling the mutated base $W_{mut}$
    \begin{enumerate}
        \item Insert $aa^{-1}$ at index 1: $a\mathbf{aa^{-1}}b$
        \item Insert $bb^{-1}$ at index 4: $aaa^{-1}b\mathbf{bb^{-1}}$
    \end{enumerate}
\end{itemize}

\textbf{Resulting Triplet}
\begin{align*}
    A &= a^{-1} a a b b^{-1} b a^{-1} \\
    P &= a b b^{-1} b a^{-1} a a^{-1} \\
    N &= a a a^{-1} b b b^{-1}
\end{align*}

Despite sharing identical string lengths and an identical degree of structural obfuscation ($N_{ops} = 2$), $A$ and $P$ mathematically reduce to the equivalence class $aba^{-1}$, while $N$ reduces to the inequivalent class $ab$.
\end{example}

\subsubsection{Geodesic length prediction}\label{sec:groups_words}

To evaluate the geodesic length estimators, a new batch of 32768 target-label tuples $(W_{\text{tangled}}, L_{\text{true}}, L_{\text{input}})$ is generated each epoch to represent the unreduced graph input, the ground-truth geodesic length, and the initial string length, respectively. The process begins by uniformly sampling an inflated random sequence of generators.

The reduction to a base representative $W_{\text{base}}$ is handled differently depending on the target group's complexity. For the Baumslag-Solitar group $BS(1,2)$, exact symbolic solvers like GAP trigger exponential memory blowup via quotient rewriting (e.g., replacing $ab \to b^2a$). To bypass this, $BS(1,2)$ sequences are reduced using a greedy, length-decreasing Dehn-style heuristic. Substrings matching more than half of a symmetrized relator are strictly replaced by their shorter inverse. Combined with memorization caching, this forces rapid, polynomial-time convergence to a highly accurate local minimum (serving as an upper bound on the true geodesic).

Conversely, for Artin Groups, the exact mathematical geodesic is computed. Because the relations consist of commuting generators, combining free reduction with continuous lexicographical sorting forms a complete and confluent rewriting system (\cite{Brieskorn1972}, \cite{Hermiller1995}). This guarantees that $W_{\text{base}}$ reaches its unique canonical normal form and absolute minimum length.

Once $W_{\text{base}}$ is computed, its length $L_{\text{true}} = \vert{}W_{\text{base}}\vert{}$ is recorded. To ensure compatibility with the classifier's output dimensions, any sequence where $L_{\text{true}} \notin [2, L_{\text{max}}]$ is discarded. Finally, to force the network to navigate the underlying geometry rather than reading pre-reduced strings, $W_{\text{base}}$ is structurally obfuscated into $W_{\text{tangled}}$ by applying $k$ random algebraic operations (e.g., inserting trivial pairs or swapping commuting generators), with the tangling volume dynamically scaled as $k = \min(\text{intensity}, \max(2, \lfloor L_{\text{true}} / 5 \rfloor))$.

\subsubsection{Wagner-Magyarik attack}

To train the metric learning architecture, batches of 32768 contrastive triplets $(A, P, N)$ are dynamically generated using a bipartite distribution. To prepare the network for cryptanalysis, 50\% of the dataset simulates a Wagner-Magyarik public-key distribution: a random identity pad is generated and combined with a secret public word to form a structurally obfuscated ciphertext $C$. The positive sample $P$ is constructed by appending the correct inverse key to $C$ and tangling the result (which mathematically reduces to the identity), while the negative sample $N$ applies an incorrect inverse key, leaving an un-canceled algebraic residual against an identity anchor $A$. The remaining 50\% of the dataset enforces general algebraic equivalence. A base word of random length is sampled from predefined length buckets utilizing either the Baumslag-Solitar generators $\{a, b, a^{-1}, b^{-1}\}$ or a localized subset of the Artin group generators. For these general triplets, the anchor $A$ is the base word, the positive $P$ is a structurally tangled variant, and the negative $N$ is a tangled version of an isometrically mutated base word. Across both distributions, the volume of applied algebraic tangling operations scales dynamically with the sequence length to maintain consistent obfuscation density across varying word lengths.

\subsection{WPNet Architecture}

Despite addressing three distinct algebraic tasks, such as Geodesic Length Estimation in $BS(1,2)$, Wagner-Magyarik (WM) Cryptanalysis, the core neural architecture relies on a shared, modular framework. The foundation is a temporal graph built from unreduced word chunks, processed by a sequence encoder (Bi-LSTM, \cite{hochreiter1997long}), refined via a Gated Graph Isomorphism Network (GINEConv, \cite{Xu2019, hu2020pretraining}), and collapsed using Attentional Aggregation. To prevent redundancy, the shared baseline is defined first, followed by the specific structural variations required for each downstream task. 

\subsubsection{Shared Core Framework}

The input to all models is an unreduced algebraic word represented as a directed graph, as described in detail in Subsection \ref{sec:sub_graph}

\paragraph{1. Gated Message Passing (AlgebraicBlock):}
Given initial continuous node embeddings $X^{(0)} \in \mathbb{R}^{N \times d}$ and projected edge attributes $E \in \mathbb{R}^{M \times d}$, the graph undergoes $K$ rounds of spatial message passing. Each layer computes a topological message $M^{(k)}$ using a GINE framework:
\begin{equation}
    M^{(k)} = \text{MLP}_{\text{gine}} \left( X^{(k)} + \text{Aggregate}(X^{(k)}, E) \right)
\end{equation}
To manage non-commutative structural data and mitigate over-smoothing, node states are updated using a learned gating matrix $W_{\text{gate}} \in \mathbb{R}^{d \times 2d}$:
\begin{equation}
    G^{(k)} = \sigma \left( W_{\text{gate}} [X^{(k)} \parallel M^{(k)}] \right)
\end{equation}
\begin{equation}
    X^{(k+1)} = \text{LayerNorm} \left( X^{(k)} + G^{(k)} \odot M^{(k)} \right)
\end{equation}

\paragraph{2. Attentional Aggregation:}
To extract a fixed-size representation of the entire word, an attentional aggregation layer (\cite{Li2016}) pools the node features. Two parallel multi-layer preceptrons (MLPs) compute a normalized scalar attention score and a feature transformation for the pooled representation:
\begin{equation}
    \alpha_i = \text{softmax} \left( \text{MLP}_{\text{gate}}(X_i) \right)
\end{equation}
\begin{equation}
    Z_{\text{pool}} = \sum_{i=1}^N \alpha_i \cdot \text{MLP}_{\text{feat}}(X_i)
\end{equation}

\subsubsection{Task-Specific Variations}

\paragraph{Geodesic Estimation:}
The geodesic model takes continuous $7$-dimensional node encodings and processes them via the Bi-LSTM. It strictly relies on the final state of the GINE layers $X^{(K)}$ for pooling. The pooled output $Z_{\text{pool}}$ bypasses normalization and is fed into a classification head (two linear layers with GELU and 20\% Dropout) to output unnormalized logits $\hat{Y} \in \mathbb{R}^{L_{\max} + 1}$.

\paragraph{Wagner-Magyarik cryptanalysis:}
The Wagner-Magyarik distinguisher model utilizes multi-scale jumping knowledge to capture both localized reductions and deep conjugate structures. Before attentional pooling, it sums the outputs across all message passing layers: $X_{\text{sum}} = \sum_{k=0}^{K} X^{(k)}$. The final pooled vector $Z_{\text{pool}}$ is $L_2$-normalized to project the embeddings onto a unit hypersphere, trained via InfoNCE to group algebraically equivalent words.

Because $A(\Gamma)$ is defined over a large generating set ($|S|=101$), continuous sinusoidal encodings lack distinction. Instead, individual letters map to a discrete embedding matrix. Each generator letter is assigned a unique positive integer index. The padding element maps to 0. The 101 positive generators map to indices 1 through 101. Their formal inverses map to indices 102 through 202. This creates a fixed vocabulary size of $V = 202$. The network initializes a trainable weight matrix of size $202 \times d$, where $d$ is the hidden dimension size (e.g., 512). Each integer ID in the 5-letter subword acts as a direct row index to retrieve a specific, independently learned $d$-dimensional vector from this matrix. The $5$ token embeddings within a subword are flattened into a wide vector ($5d$) before Bi-LSTM processing. Following the message-passing layers, a single learned parameter tensor $Z_{\text{global}}$ is added identically to all $N$ nodes ($\tilde{X}^{(K)}_i = X^{(K)}_i + Z_{\text{global}}$) to anchor local features to a shared baseline before pooling and $L_2$-normalization. Local message-passing layers struggle to propagate information across extremely long sequences. Adding a global tensor to every node allows long-range, graph-wide structural properties to be accessed instantly across the entire graph without getting lost in deep message-passing steps. In tasks relying on triplet loss and similarity comparisons (like the Wagner-Magyarik cryptographic attack), the global tensor anchors the final pooled embedding to a shared, learned coordinate space, improving the separation margin between valid and corrupted words.

Table \ref{tab:architecture_variations} details the specific structural variations applied to this shared core for the three distinct experimental tasks.

\begin{table}[h!]
    \centering
    \caption{Architectural variations across the three algebraic tasks, where $d$ is the hidden dimension ($256$ to $512$), $d_{\text{out}}$ is the output dimension, $N$ is the number of nodes, and $L_{\max}$ is the maximum exact length limit.}
    \footnotesize
    \vspace{0.5em}
    \begin{tabular}{p{0.18\textwidth} p{0.25\textwidth} p{0.25\textwidth} p{0.25\textwidth}}
        \toprule
        \textbf{Component} & \textbf{Geodesic Estimation ($BS(1,2)$)} & \textbf{WM Metric Attack ($BS(1,2)$)} & \textbf{Artin Embedding} \\
        \midrule
        \textbf{Node Encoding} & Continuous sinusoidal ($7$-dim) & Continuous sinusoidal ($7$-dim) & Discrete ID lookup ($V=201$) \\
        \textbf{Chunk Size} & $C=5$ & $C=6$ & $C=5$ (flattened sequence) \\
        \textbf{Bi-LSTM Input} & Node level ($N \times 7$) & Node level ($N \times 7$) & Token level ($N \times 5d$) \\
        \textbf{Integration} & Terminal output only ($X^{(K)}$) & Multi-scale summation ($\sum X^{(k)}$) & Global token shift ($+ Z_{\text{global}}$) \\
        \textbf{Output Head} & Linear classifier (logits) & $L_2$ Normalization & $L_2$ Normalization \\
        \textbf{Objective} & Cross-Entropy (Length Class) & InfoNCE (Cosine Similarity) & InfoNCE (Cosine Similarity) \\
        \textbf{Final Dim.} & $L_{\max} + 1$ classes & $d_{\text{out}} = 128$ hypersphere & $d_{\text{out}} = 128$ hypersphere \\
        \bottomrule
    \end{tabular}
    \label{tab:architecture_variations}
\end{table}

\subsection{Training Procedure}

For the task of embedding words in the hypersphere, the InfoNCE objective \cite{oord2018representation, chen2020simple} uses in-batch negative sampling to contrast an anchor $A_j$ against its hard positive $P_j$, hard negative $N_j$, and all other non-matching batch elements. The loss is defined as:$$\mathcal{L} = -\frac{1}{B} \sum_{j=1}^{B} \log \frac{\exp(\text{sim}(A_j, P_j) / \tau)}{\exp(\text{sim}(A_j, P_j) / \tau) + \sum_{k=1}^{B} \exp(\text{sim}(A_j, N_k) / \tau) + \sum_{k \neq j} \exp(\text{sim}(A_j, P_k) / \tau)}$$where $\tau$ is the temperature and $\text{sim}(u, v)$ is the cosine similarity between $L_2$-normalized representations. To maintain a stable identity reference vector, elements reducing to the trivial identity are explicitly flagged. A False Negative Mask prevents the InfoNCE loss from repelling these equivalent elements, while an Identity Gravity Penalty actively pulls them toward their batch centroid.Optimization is executed in PyTorch \cite{paszke2019pytorch} via AdamW \cite{loshchilov2017decoupled} (learning rate $10^{-4}$, weight decay $10^{-4}$), gradient clipping (max norm $1.0$) \cite{pascanu2013difficulty}, and a plateau scheduler (factor $0.5$, patience $5$).

For geodesic length classification, the network is optimized using cross-entropy loss (\cite{Shannon1948, Goodfellow2016}). Cross-entropy measures the divergence between the model's predicted probability distribution over discrete length classes and the true discrete length labels. By treating each possible word length as an independent category rather than assuming a continuous scalar progression, cross-entropy forces the network to construct sharp, highly specific decision boundaries for exact integer lengths, avoiding the outlier sensitivity and averaging issues typical of regression.

For the Wagner-Magyarik attack, optimization employs Triplet Margin Loss \cite{schroff2015facenet, weinberger2009distance}. This objective function structures the latent representation space by evaluating triplets consisting of an anchor ($a$), a positive sample ($p$), and a negative sample ($n$). The loss penalizes the model unless the distance between the anchor and the positive is strictly smaller than the distance between the anchor and the negative by at least a predefined margin $m$. Expressed mathematically as:$$\mathcal{L}(a, p, n) = \max \{ d(a, p) - d(a, n) + m, 0 \}$$this formulation ensures tight clustering of algebraically equivalent elements while enforcing a strict geometric boundary against non-equivalent cryptographic mutations.

\subsection{Model evaluation}

The model is evaluated on three tasks. The first task is similar to the model training setup, where triplets of words are presented and the model's embeddings are tested. In the second task, the models must predict the geodesic length of an unreduced word in $BS(1,2)$ and the Artin group $A(\Gamma)$. The third task is the Wagner-Magyarik cryptosystem attack.

\subsubsection{Word embeddings}

The model's distinguishing capability is evaluated using a triplet-based metric learning framework. For each test, the algorithm generates a triplet consisting of an anchor word, a positive word, and a negative word based on the following specific criteria: The positive word is a geodesic representative that is algebraically equivalent to the anchor word. The negative word is a group element constructed by randomly scrambling 25\% to 50\% of the letters from the positive word. The network projects the corresponding graphs into a normalized embedding space to compute the cosine similarities for the anchor-positive pair ($s_{pos}$) and the anchor-negative pair ($s_{neg}$). A prediction is classified as correct only if the positive similarity exceeds the negative similarity by a strict margin threshold of 0.2 ($s_{pos} - s_{neg} > 0.2$). Final accuracy percentages and margin statistics are then aggregated across the entire test suite.

\subsubsection{Geodesic length prediction}

To evaluate the geodesic length estimator, test datasets are systematically generated and stratified by their true reduced length. The evaluation domain, spanning up to the classification limit $L_{\max} = 200$, is partitioned into ten uniformly sized length buckets. For each bucket, exactly 100 valid test samples are generated. The generation process begins by constructing a random sequence of generators with an initial length uniformly sampled up to four times the bucket's maximum boundary. This heavily inflated sequence is reduced to establish its ground-truth geodesic length, and discarded if it falls outside the targeted bucket. Note that the samples generated here follow the same strategy as the training data generation and might not be the true geodesic representative in $BS(1,2)$. In the Artin group the reduction algorithm guarantees the true geodesic representative.

To guarantee the final input remains strictly non-geodesic, the reduced core undergoes a final tangling phase where a dynamically scaled number of random algebraic operations (inserting trivial pairs or symmetrized relators, bounded to a maximum of 3 operations) is applied. Model performance is then quantitatively assessed using exact match accuracy, the strict percentage of instances where the network's predicted length class perfectly matches the true length and the Mean Absolute Error (MAE) of the predictions.

\subsubsection{Wagner-Magyarik Attack}

To evaluate the network's zero-shot algebraic understanding, the model is periodically suspended during training and applied as an equality oracle in two distinct cryptographic attacks. Both tests rely on an \textit{identity reference vector}, computed by passing eight independently tangled trivial identity words through the network and averaging their $L_2$-normalized latent embeddings. 

\textbf{Test 1: Wagner-Magyarik Direct Decryption}
\begin{enumerate}
    \item \textbf{Generation:} The environment generates two random public words ($P_0, P_1$) of length 5 and selects a secret message bit $b \in \{0, 1\}$. A ciphertext is constructed as $C = R \cdot P_b$, where $R$ is a highly tangled identity pad (generated with noise length $N_{len}=10$).
    \item \textbf{Model Application:} The attacker constructs two hypothesis strings: $T_0 = C \cdot P_0^{-1}$ and $T_1 = C \cdot P_1^{-1}$. The model encodes $T_0$ and $T_1$ into the latent space. The network computes the cosine similarity of both embeddings against a stabilized identity reference vector (the computed centroid of 64 independently generated cryptographic identity pads). 
    \item \textbf{Accuracy Calculation:} The model predicts bit $b=0$ if $\text{sim}(T_0, \text{Id}) > \text{sim}(T_1, \text{Id})$, and $b=1$ otherwise. A trial is marked successful if the prediction matches the true secret bit. Total accuracy is the percentage of correct predictions over 50 independent trials.
\end{enumerate}

\begin{example} Wagner-Magyarik Direct Decryption:

Consider the Baumslag group $G_{1,2} = \langle a, b \mid b a b^{-1} = a^2 \rangle$. The generators are mapped numerically as $a=1, b=2, a^{-1}=-1, b^{-1}=-2$.

\textbf{1. Initialization and Encryption:}
\begin{itemize}
    \item \textbf{Public Words:} $P_0 = [1, 2, 1, -2, 1]$ (representing $a b a b^{-1} a$) and $P_1 = [2, -1, 2, 1, 1]$ (representing $b a^{-1} b a^2$).
    \item \textbf{Secret Bit:} $b = 0$, making the target word $P_0$.
    \item \textbf{Identity Pad ($R$):} A highly tangled string generated via random insertions of $a a^{-1}$, $b b^{-1}$, and the relator $b a b^{-1} a^{-2}$. Mathematically, $R$ reduces to the empty word.
    \item \textbf{Ciphertext ($C$):} $C = R \cdot P_0$.
\end{itemize}

\textbf{2. Hypothesis Construction:}
Two test strings are formed by appending the inverse of each public word to the ciphertext:
\begin{align*}
    T_0 &= C \cdot P_0^{-1} = (R \cdot P_0) \cdot P_0^{-1} = R \\
    T_1 &= C \cdot P_1^{-1} = (R \cdot P_0) \cdot P_1^{-1}
\end{align*}
Mathematically, $T_0$ simplifies exactly to the identity pad $R$, while $T_1$ remains a complex, non-trivial group element.

\textbf{3. Evaluation:}
The network embeds both $T_0$ and $T_1$ into the latent space and computes their cosine similarities against the stabilized identity centroid ($\text{Id}$):
\begin{align*}
    \text{sim}(T_0, \text{Id}) &= 0.94 \quad \text{(High structural similarity to identity)} \\
    \text{sim}(T_1, \text{Id}) &= 0.12 \quad \text{(Recognized as a non-identity element)}
\end{align*}

\textbf{4. Conclusion:}
Since $\text{sim}(T_0, \text{Id}) > \text{sim}(T_1, \text{Id})$, the network correctly predicts the secret bit $b=0$. The decryption trial is marked successful.

\end{example}

\subsection{Computational Methods}

All code was developed in Python 3.11.2 using Pytorch (2.10.0+cu128) on an Nvidia Quadro RTX 8000 or alternatively Python 3.13.9 and Pytorch (2.14.0.dev20260720+cu130) when working with the Nvidia Geforce RTX 5080.

\section{Results and Discussion}

The results are structured into three distinct parts, covering both the Baumslag-Solitar group $BS(1,2)$ and the Artin group $A(\Gamma)$. The first part analyzes the embedding properties of equivalent and non-equivalent words. The second part presents the performance of the geodesic length prediction models, and the final part evaluates the cryptanalytic attack on the Wagner-Magyarik protocol.

\subsection{Word embeddings}

The model embeds words in each group into the unit sphere in $\mathbb{R}^{128}$. When an unreduced word $w$ is represented by a geodesic segment $g$, the embeddings are close to each other, in other words, they have a high cosine similarity. When a geodesic segment $h$ is similar to $g$ and hence $w$, but is not equal, then the cosine similarity of the embeddings is low. The models are trained on triplets $(w, p,n)$, where $w$ is an unreduced word, $p$ a reduced representative and $n$ a reduced word that is similar to $p$ but not equal.

Before evaluating the distinguisher against the cryptographic protocol, the topological validity of the learned embedding space must be established. The network must demonstrate that it maps visually distinct but algebraically equivalent words (positive pairs) to proximate coordinates, while mapping inequivalent words (negative pairs) to distant orthogonal coordinates. 

\subsubsection{Word embeddings}

The separation metric is defined as the margin $\Delta = S_{pos} - S_{neg}$, where $S$ denotes the cosine similarity in the learned metric space. To test for zero-shot algebraic generalization, the evaluation was conducted across escalating word lengths up to $L=25000$, well exceeding the examples seen in the training data where lengths were capped at $100$.

\begin{figure}[H]
\centering
\begin{subfigure}[b]{0.48\textwidth}
    \centering
    \includegraphics[width=\linewidth]{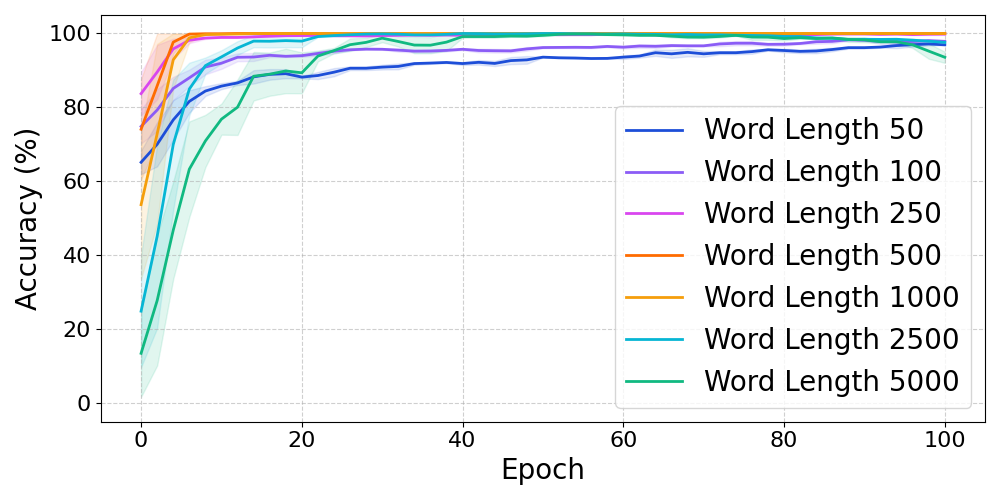}
    \caption{}
    \label{fig:bs_wp_learning_curve}
\end{subfigure}
\hfill
\begin{subfigure}[b]{0.48\textwidth}
    \centering
    \includegraphics[width=\linewidth]{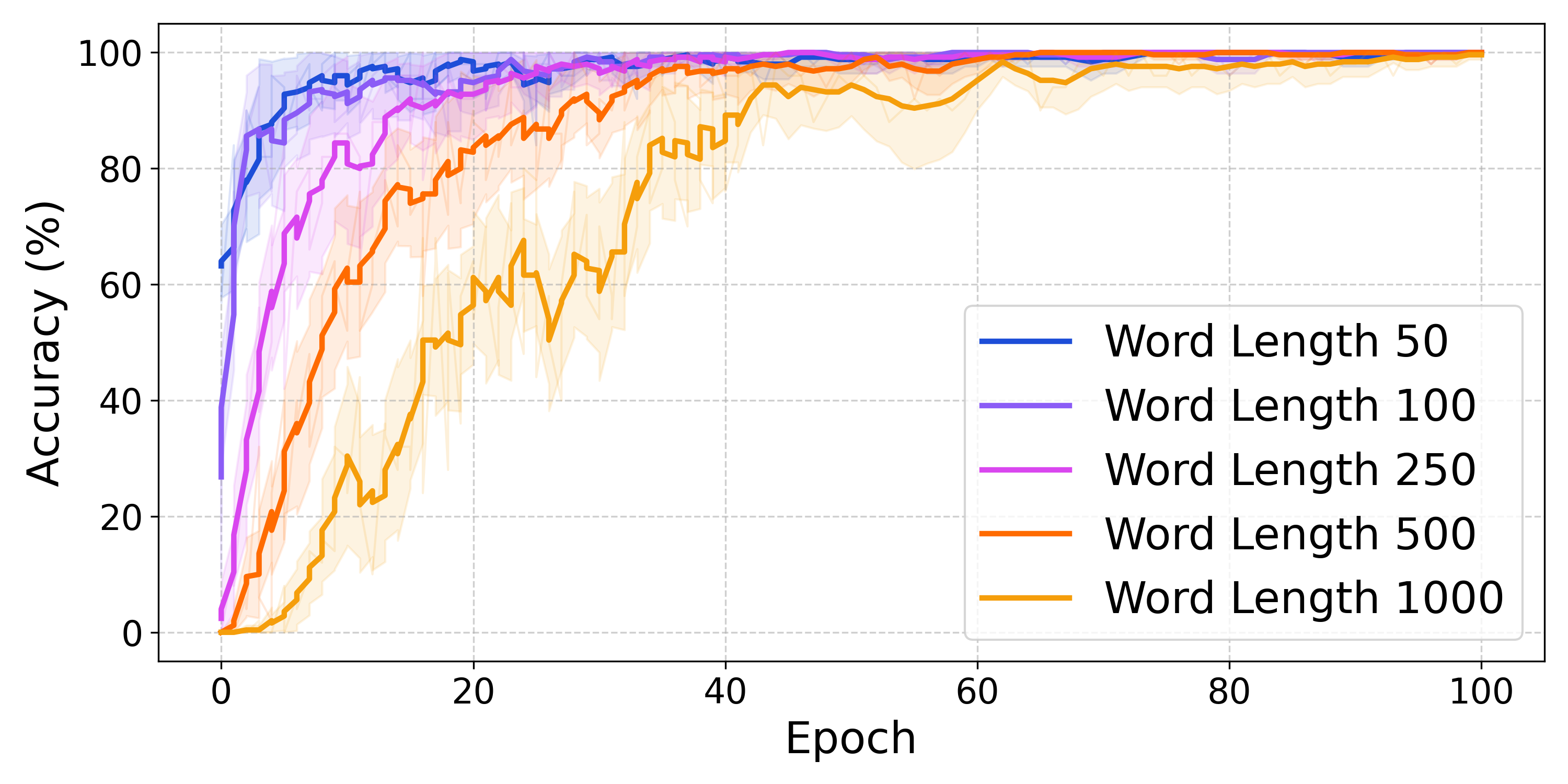}
    \caption{}
    \label{fig:artin_wp_learning_curve}
\end{subfigure}
\caption{(a) Learning curve for word embeddings in BS(1,2). (b) Learning curve for word embeddings in $A(\Gamma)$.}
\label{fig:combined_wp_learning_curves}
\end{figure}

Analysis of the training run reveals a two-stage learning dynamic (Figure \ref{fig:bs_wp_learning_curve}). In the initial phase, the architecture rapidly generalizes, solving the word problem with 100\% accuracy for sequences up to $L=250$ by the second epoch. 

At this peak convergence state, the model achieves a mean classification accuracy of 99.93\% when considering words of length up to 5000. It should be emphasized here that the model only saw words of length up to 100 in training. The generalization as demonstrated in Table \ref{tab:combined_wp_results} shows very good identification well beyond that threshold.

Notably, an inverse relationship between word length and the separation margin emerges at extreme scales. As $L$ increases beyond 250, the geometric margin compresses (decreasing from +0.599 at $L=100$ to +0.298 at $L=5000$). This continuous compression indicates that as the length of the sequence grows, the graph embeddings become denser, pulling the negative representations closer to the positive cluster. However, because the difference remains strictly positive with low standard deviation, the decision boundary is perfectly preserved. This confirms that the network has not memorized finite token patterns, but has successfully internalized the non-commutative quotient geometry of the group at an asymptotic scale, which is in particular emphasized by still achieving 74.90\% accuracy on words of length 20,000.

The task for evaluating the learning process of the Artin-WPNet was set up analogously to the task for the BS-WPNet. The success rates of both models are compared in Table \ref{tab:combined_wp_results}. For the Artin group, practical reasons required capping the word length at 1000. The Artin-WPNet was evaluated in five tests on 10 samples for each word length, generating the average success rate and standard deviation. The approximate word reduction described in Section \ref{sec:groups_words} in $BS(1,2)$ is much more efficient than in $A(\Gamma)$. Further, as will be seen in the results of the geodesic length prediction, the words in $BS(1,2)$ reduce at a much higher rate than in the Artin group. This led to extensive memory demands for longer words, rendering the investigation of word lengths 2500 and 5000 unfeasible for the Artin group. 

\begin{table}[H]
\centering
\caption{Word Embedding Classification Results for $BS(1,2)$ and $A(\Gamma)$}
\label{tab:combined_wp_results}
\footnotesize
\begin{tabular}{c cc cc}
\toprule
& \multicolumn{2}{c}{\textbf{BS(1,2)}} & \multicolumn{2}{c}{\textbf{Artin Group}} \\
\cmidrule(lr){2-3} \cmidrule(lr){4-5}
\textbf{Word Length} & \textbf{Accuracy (\%)} & \textbf{Margin ($\pm$ Std)} & \textbf{Accuracy (\%)} & \textbf{Margin ($\pm$ Std)} \\
\midrule
50 & 97.00\% $\pm$ 1.58\% & +0.565 $\pm$ 0.024 & 100.00\% $\pm$ 0.00\% & +0.744 $\pm$ 0.055 \\
100 & 97.60\% $\pm$ 1.28\% & +0.599 $\pm$ 0.016 & 100.00\% $\pm$ 0.00\% & +0.585 $\pm$ 0.027 \\
250 & 100.00\% $\pm$ 0.00\% & +0.537 $\pm$ 0.010 & 100.00\% $\pm$ 0.00\% & +0.471 $\pm$ 0.031 \\
500 & 100.00\% $\pm$ 0.00\% & +0.428 $\pm$ 0.016 & 98.00\% $\pm$ 4.00\% & +0.356 $\pm$ 0.022 \\
1000 & 100.00\% $\pm$ 0.00\% & +0.347 $\pm$ 0.020 & 96.00\% $\pm$ 4.90\% & +0.310 $\pm$ 0.009 \\
2500 & 97.50\% $\pm$ 0.71\% & +0.328 $\pm$ 0.022 & -- & -- \\
5000 & 94.50\% $\pm$ 1.58\% & +0.298 $\pm$ 0.016 & -- & -- \\
10,000 & 86.10\% $\pm$ 1.50\% & +0.224 $\pm$ 0.019 & -- & -- \\
15,000 & 79.90\% $\pm$ 2.48\% & +0.191 $\pm$ 0.018 & -- & -- \\
20,000 & 74.90\% $\pm$ 1.77\% & +0.179 $\pm$ 0.015 & -- & -- \\
\bottomrule
\end{tabular}
\end{table}

The final classification results of Artin-WPNet in Table \ref{tab:combined_wp_results} are very similar to the ones of $BS(1,2)$. The larger standard deviation stems from the lower number of tests that were possible. The success rate per word length and epoch is shown in Figure \ref{fig:artin_wp_learning_curve}. It can be seen that in contrast to the results in Figure \ref{fig:bs_wp_learning_curve} for $BS(1,2)$, the Artin-WPNet takes a significantly longer time to learn. Further, each epoch took on average $35.132 \pm 2.946$ seconds for training data generation, $9.308 \pm 2.668$ seconds for gradient descent and around another $2 \text{ min } 38 \text{ sec}$ ($\pm 1 \text{ min } 22 \text{ sec}$) for evaluation, significantly longer than BS-WPNet (9.07s $\pm$ 0.58s for data generation, 7.21s $\pm$ 1.09s for each epoch training and 8.11s $\pm$ 0.44s for testing) due to the more complex data generation.

It can also be seen in Figure \ref{fig:artin_wp_learning_curve} that the WPNet learns to correctly classify shorter words first, and words with noise of length up to 1000 take almost 80 epochs to reach a classification accuracy of at least 95\%. The learning curves in Figure \ref{fig:artin_wp_learning_curve} also show a much higher variability than the learning curves for $BS(1,2)$ in Figure \ref{fig:bs_wp_learning_curve}. This is suspected to be attributed to the increased number of generators and the overall algebraic complexity.

Generating and testing 1000 words of length 1000 in $BS(1,2)$ took $14.362 \pm 0.316s$ seconds for generation and  $0.181 \pm 0.042$ seconds for inference. 
In the Artin group $A(\Gamma)$ data generation and testing for 1000 words of length 1000 in the Artin group took on average $135.297 \pm 20.301$ seconds for generation and $1.447 \pm 0.054$ seconds for inference. It can already be seen that inference is significantly faster than even test data generation, further highlighting that WPNet can act as a fast approximate solution to the word problem.

\subsection{Geodesic length prediction}

\begin{figure}[htb]
\centering

% Figure (a)
\begin{subfigure}[t]{0.48\textwidth}
    \centering
    \includegraphics[width=\linewidth]{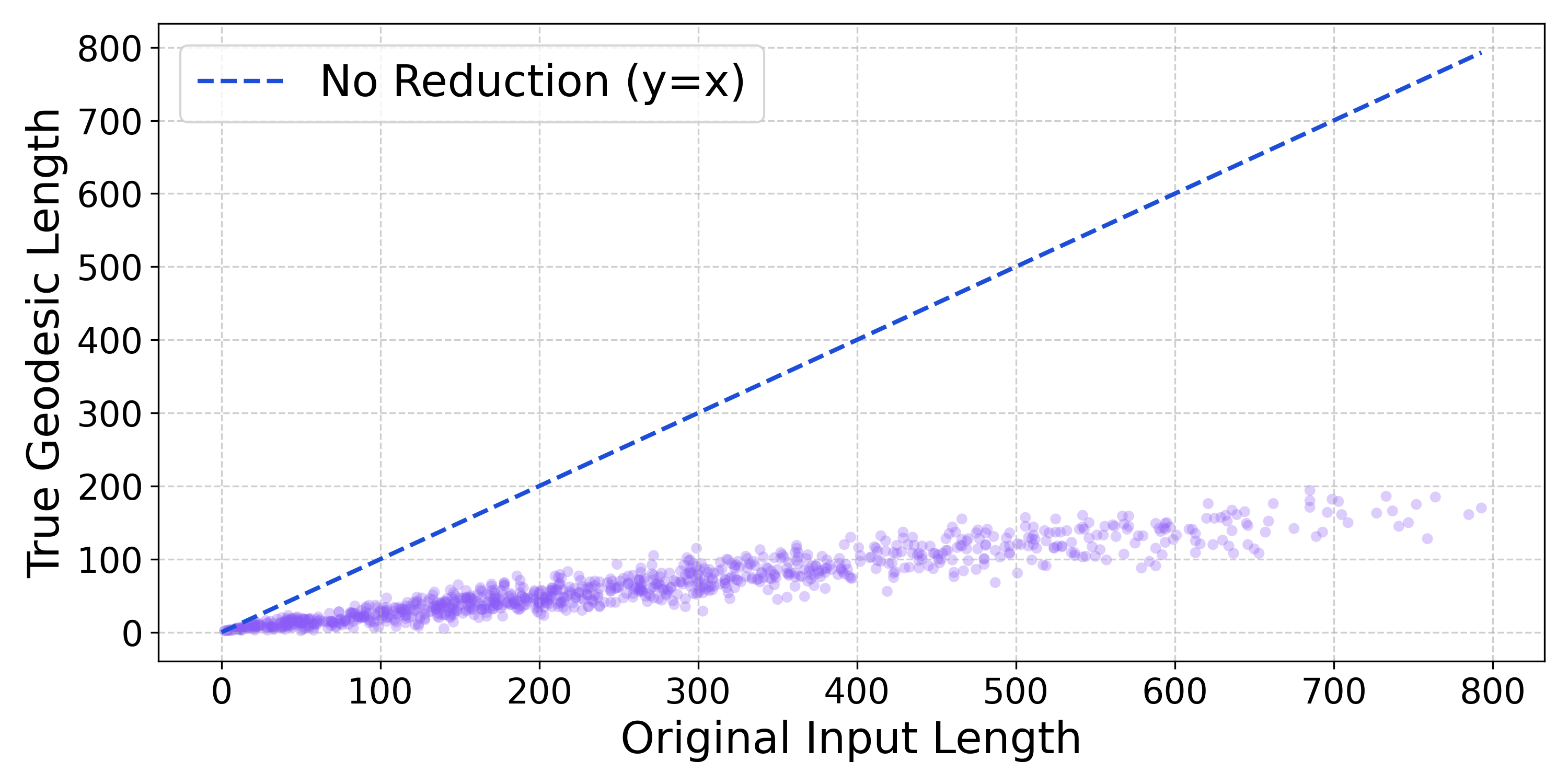}
    \caption{}
    \label{fig:BS_prediction_analysis}
\end{subfigure}
\hfill
% Figure (b)
\begin{subfigure}[t]{0.48\textwidth}
    \centering
    \includegraphics[width=\linewidth]{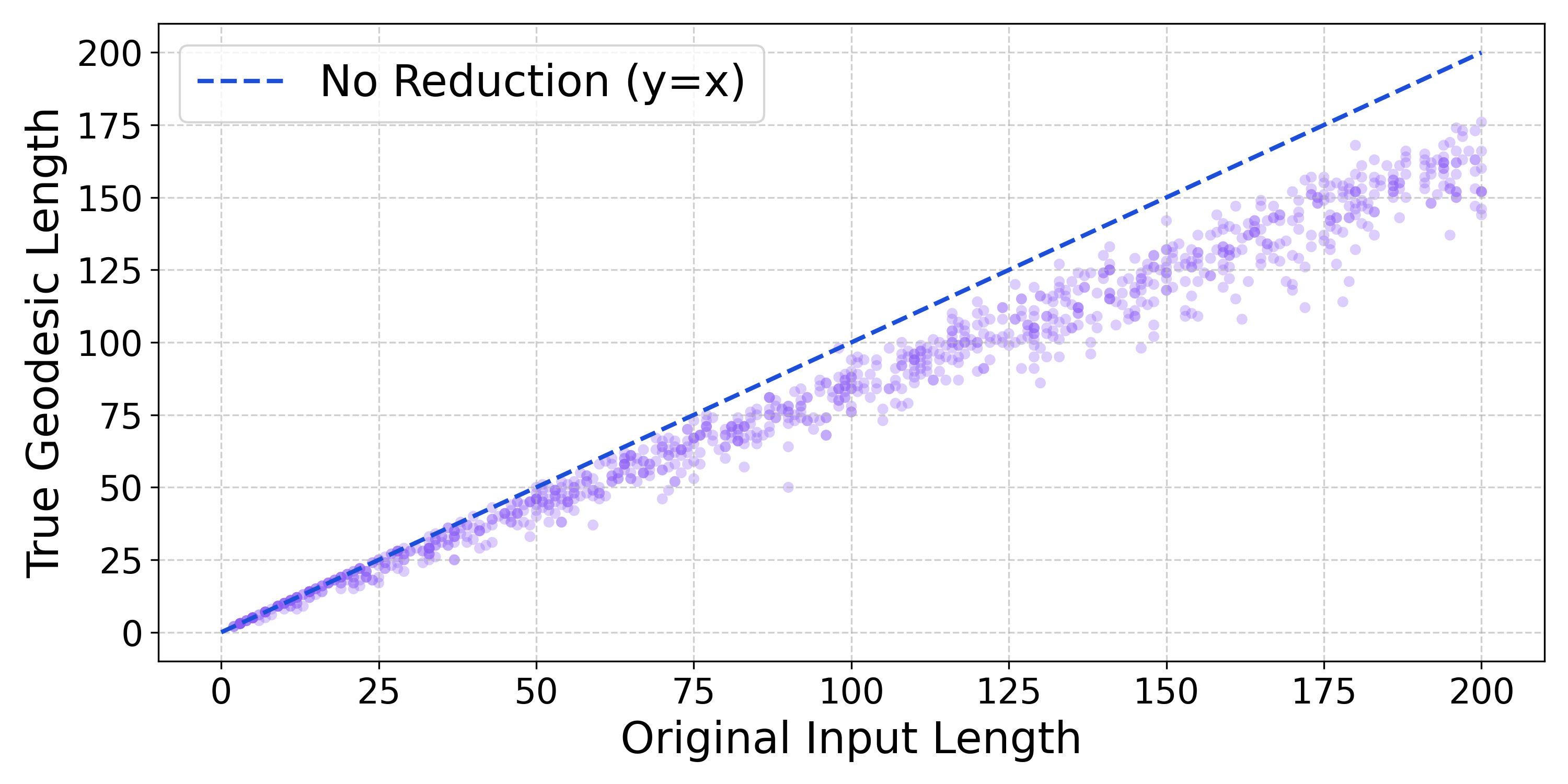}
    \caption{}
    \label{fig:artin_prediction_analysis}
\end{subfigure}

\vspace{0.5cm} % Vertical space between rows

% Figure (c)
\begin{subfigure}[t]{0.48\textwidth}
    \centering
    \includegraphics[width=\linewidth]{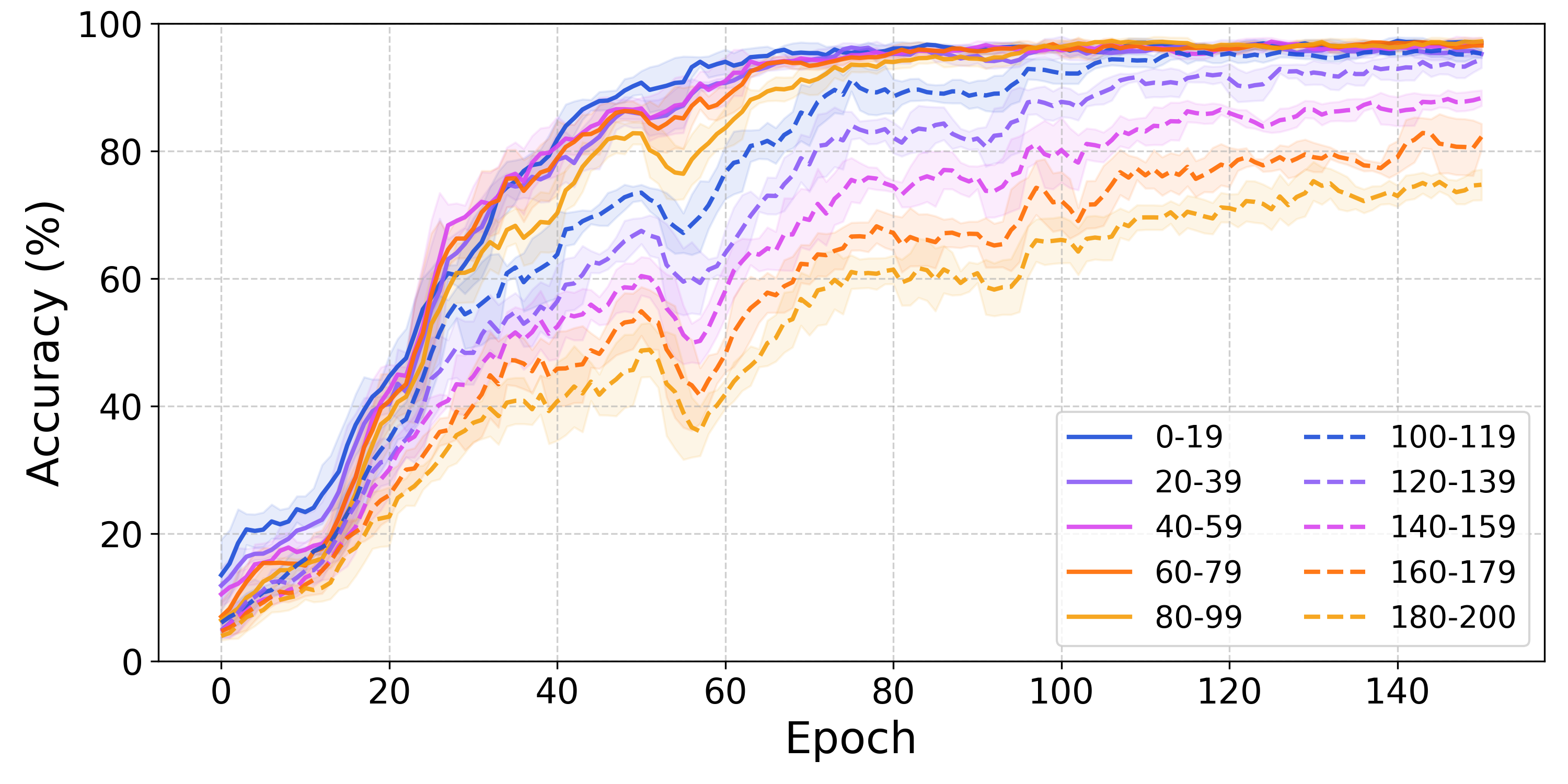}
    \caption{}
    \label{fig:bs_geo_bucket}
\end{subfigure}
\hfill
% Figure (d)
\begin{subfigure}[t]{0.48\textwidth}
    \centering
    \includegraphics[width=\linewidth]{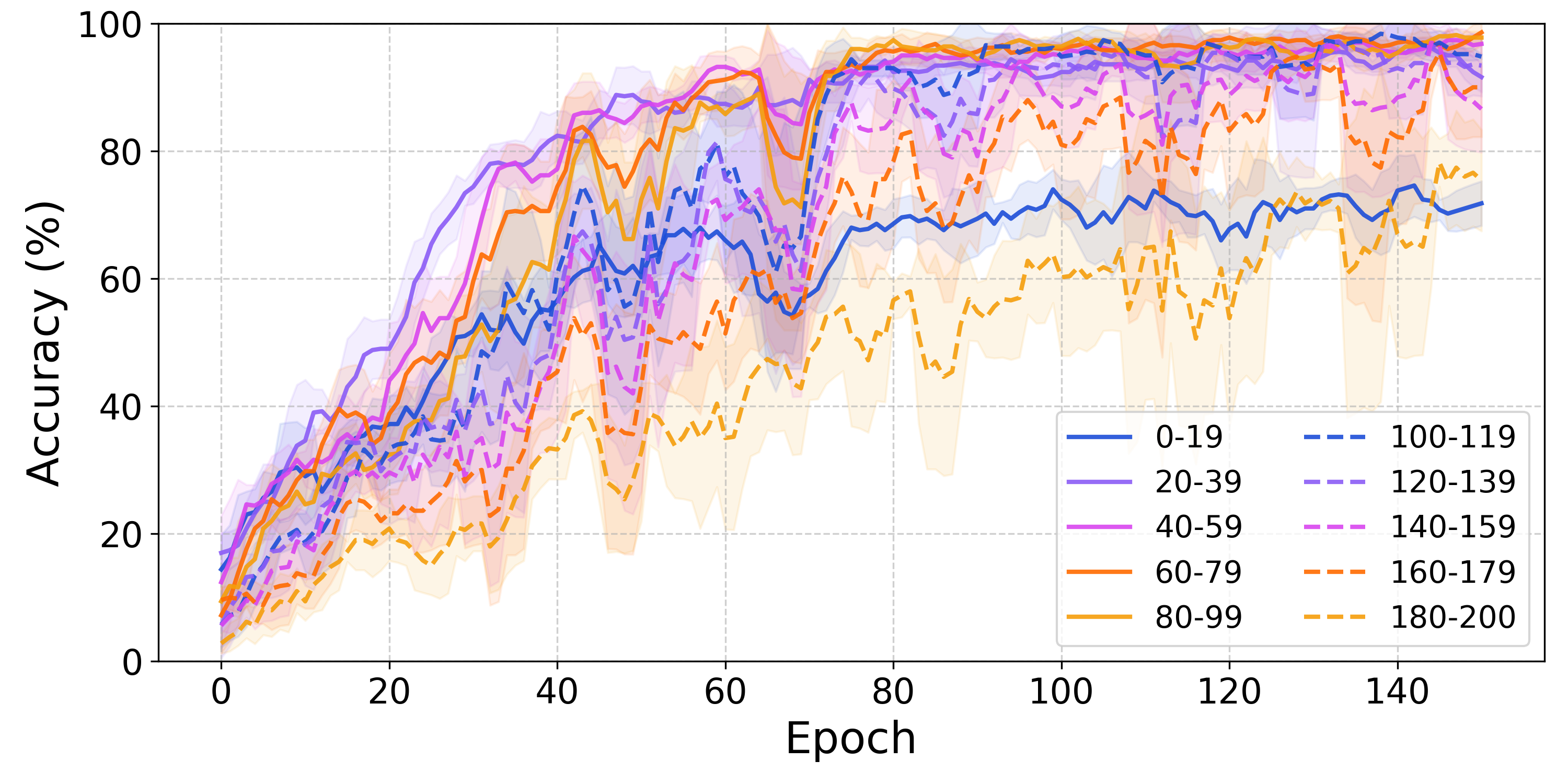}
    \caption{}
    \label{fig:Artin_geo_bucket}
\end{subfigure}
\caption{Word length reduction in $BS(1,2)$ (a) and $A(\Gamma)$ (b). The training progress for the accuracies per length bucket are shown in (c) for $BS(1,2)$ and in (d) for $A(\Gamma)$.}
\label{fig:combined_analysis}
\end{figure}

The metabelian group $BS(1,2)$ heavily compresses unreduced words. Figure \ref{fig:BS_prediction_analysis} demonstrates this, showing true geodesic representations are typically 10\% to 20\% of the original input length. Analyzing the unreduced versus geodesic lengths in $A(\Gamma)$ reveals that the shrinkage is less pronounced than in $BS(1,2)$, as illustrated in Figure \ref{fig:artin_prediction_analysis}. The parameters for the random Artin group ($p=0.9$) were chosen aggressively to yield a shrinkage of approximately $20\%$.

The WPNet for $BS(1,2)$ and $A(\Gamma)$ further exhibit slightly different learning curves, as shown in Figure \ref{fig:bs_geo_bucket} and \ref{fig:Artin_geo_bucket}.  Configured as a 200-class classifier, WPNet evaluated these lengths over 150 epochs. 

Table \ref{tab:geo_combined_performance} outlines the peak results for both groups, highlighting a strong global accuracy of 80.30\% (MAE = 0.26) for $BS(1,2)$. The model excels on shorter words ($> 90\%$ accuracy) but gradually declines for geodesics exceeding 120 letters (derived from unreduced words up to length 800). Figure \ref{fig:bs_geo_bucket} illustrates uniform convergence, with longer words consistently trailing shorter sequences.

Crucially, while exact match accuracy decreases for longer sequences (e.g., 57\% for lengths 180-200), the MAE remains well below 1 across all buckets. This sub-unit error indicates that even incorrect length predictions deviate from the true geodesic length by merely one or two letters. Unsurprisingly, since the WPNet has been trained as a classifier, the generalization above length 200 is marginal, with error rates of 97\% at lengths 200-300 and 0\% accuracy above that.

Table \ref{tab:geo_combined_performance} details the models' performance across length buckets at peak convergence. The model demonstrates exceptional precision on mid-length sequences (30 to 179 letters), consistently achieving exact match accuracies above $88\%$ with a Mean Absolute Error (MAE) under $0.3$.

\begin{table}[H]
\centering
\caption{Performance Breakdown by Length Bucket at Peak Convergence for $BS(1,2)$ and $A(\Gamma)$}
\label{tab:geo_combined_performance}
\footnotesize
\begin{tabular}{l c c c c c c c}
\toprule
& \multicolumn{3}{c}{\textbf{BS(1,2)}} & \multicolumn{4}{c}{\textbf{Artin Group}} \\
\cmidrule(lr){2-4} \cmidrule(lr){5-8}
\textbf{Length} & \textbf{Avg. Input} & \textbf{MAE} & \textbf{Acc. (\%)} & \textbf{Avg. Input} & \textbf{Avg. Red.} & \textbf{MAE} & \textbf{Acc. (\%)} \\
\midrule
0-19 & 33.8 & 0.14 & 95.00\% & 10.6 & 9.9 & 3.18 & 73.00\% \\
20-39 & 79.6 & 0.05 & 98.00\% & 29.2 & 26.3 & 0.20 & 90.00\% \\
40-59 & 127.7 & 0.17 & 91.00\% & 49.5 & 43.7 & 0.08 & 96.00\% \\
60-79 & 184.0 & 0.16 & 93.00\% & 69.3 & 60.0 & 0.00 & 100.00\% \\
80-99 & 239.6 & 0.18 & 86.00\% & 89.2 & 75.0 & 0.06 & 97.00\% \\
100-119 & 266.8 & 0.27 & 76.00\% & 109.9 & 92.4 & 0.10 & 95.00\% \\
120-139 & 346.2 & 0.39 & 67.00\% & 129.1 & 107.2 & 0.12 & 94.00\% \\
140-159 & 379.6 & 0.25 & 76.00\% & 150.1 & 123.1 & 0.20 & 90.00\% \\
160-179 & 423.6 & 0.46 & 64.00\% & 169.7 & 138.9 & 0.08 & 96.00\% \\
180-200 & 500.9 & 0.57 & 57.00\% & 189.9 & 155.8 & 0.52 & 76.00\% \\
\bottomrule
\end{tabular}
\end{table}

Performance degrades at both extremes: the shortest sequences (0 to 19 letters) struggle with a high MAE of 8.79, while the longest sequences (180 to 200 letters) show a sharp decline in accuracy to $54\%$ as the MAE exceeds 1.0. Similarly to the geodesic length prediction for $BS(1,2)$, the generalization beyond word length 200 is very poor, as to be expected when using a classifier.

\subsection{Attack on Wagner-Magyarik}

In the Wagner-Magyarik test the WPNet was trained to correctly identify one of two public words $W_{p_1}, W_{p_2}$, which were obscured by words $w,v$, where $w \equiv 1, v \equiv 1$ in $G=BS(1,2)$ or $G=A(\Gamma)$. The test was set up to increase the length of $w$ and $v$ and the success rate in correctly identifying either $W_{p_1}$ or $W_{p_2}$ was recorded. Table \ref{tab:combined_WM_peak_performance} summarizes the results.

With an increased length in the noise word the certainty of the detection in $BS(1,2)$ drops. This is reflected in the margin reducing to as low as 0.085 for noise word lengths of 1000, which means that the embedding of both words lie very close to each other, but are still distinguishable in most cases, which the accuracy of 97.80\% reflects. On the other hand, the model for the Artin group $A(\Gamma)$ is well capable of separating longer words, which the margin of  1.035 at length 1000 shows.

\begin{table}[H]
\centering
\caption{Performance of WPNet for WM on $BS(1,2)$ and $A(\Gamma)$}
\footnotesize
\begin{tabular}{ccccc}
\hline
 & \multicolumn{2}{c}{\textbf{$BS(1,2)$}} & \multicolumn{2}{c}{\textbf{$A(\Gamma)$}} \\
\cline{2-3} \cline{4-5}
\textbf{Length} & \textbf{Accuracy $\pm$ std} & \textbf{Margin} & \textbf{Accuracy $\pm$ std} & \textbf{Margin} \\
\hline
10   & 99.20\% $\pm$ 0.40\% & +1.706 $\pm$ 0.435 & $99.00 \pm 2.00$ & $+1.112 \pm 0.190$ \\
50   & 98.40\% $\pm$ 0.80\% & +1.320 $\pm$ 0.647 & $98.00 \pm 2.45$ & $+1.050 \pm 0.039$ \\
100  & 97.60\% $\pm$ 1.36\% & +0.709 $\pm$ 0.573 & $98.00 \pm 4.00$ & $+0.985 \pm 0.127$ \\
500  & 97.40\% $\pm$ 1.74\% & +0.162 $\pm$ 0.151 & $98.00 \pm 2.45$ & $+1.053 \pm 0.062$ \\
1000 & 97.80\% $\pm$ 1.33\% & +0.085 $\pm$ 0.070 & $97.00 \pm 2.45$ & $+1.035 \pm 0.088$ \\
\hline
\end{tabular}
\label{tab:combined_WM_peak_performance}
\end{table}

Training just takes on average $11.37s$ $\pm$ 2.05s, with another 6.42s $\pm$ 0.49s for the data generation, per epoch. Testing 5x100 words of each length bucket took on average 2min 20.43s ($\pm$ 6.53s), with more than 50\% of the testing time allocated to the in total 500 words of length 1000. 

Data generation for the 8192 triplets in $A(\Gamma)$ required an average of 40.05s $\pm$ 17.08s per epoch. This high variance stems from the fluctuating shrink factor between the unreduced and reduced words. Longer unreduced sequences generated larger graph structures, which subsequently increased the training time to 18.12s $\pm$ 6.01s per epoch. Testing durations remained efficient at 21.87s $\pm$ 1.51s. A notable distinction emerges regarding the detection margin under increasing noise. As shown in Table \ref{tab:combined_WM_peak_performance}, the certainty margin for $BS(1,2)$ drops drastically to 0.085 at a noise length of 1000. In contrast, the margin for $A(\Gamma)$ remains highly stable at 1.035 under identical noise conditions. This indicates a surprising robustness in the generated embeddings, particularly given the notorious computational difficulty of certain Artin groups.

Ultimately, WPNet successfully uncovers the hidden messages across both groups, further confirming the established vulnerabilities of the Wagner-Magyarik protocol. While this success is expected for the computationally tractable $BS(1,2)$ group, it is highly significant for the Artin group $A(\Gamma)$, where the word problem in the presence of numerous braid relations is not generally known to be decidable.

\section{Discussion}

The results demonstrate that WPNet provides a fast and efficient heuristic approximation for the word problem. While the word problem in the Baumslag-Solitar group $BS(1,2)$ is mathematically well understood, it remains notoriously difficult, and often undecidable, for general Artin groups containing numerous braid relations. The success of the network on the latter indicates that this neuro-symbolic approach can scale effectively to other computationally hostile groups. Notably, execution runtimes are magnitudes shorter than those of exact symbolic solvers like SageMath (\cite{Holt2005, Epstein1992}), and inference times require only a fraction of the time needed for data generation.

This approach does not offer a mathematically universal solution for infinite general group structures. The outputs are strictly heuristic, and the evaluations were restricted to word lengths up to 20,000. However, this bounded computational space perfectly covers the word lengths that are practically feasible and relevant for real-world cryptographic applications.

Rather than relying on a single universal architecture, a variety of models were shaped specifically for their respective learning tasks. The flexibility of WPNet allows it to be easily adapted and rapidly retrained for other algebraic structures. Furthermore, training these specialized models proved highly efficient, typically concluding within a few hours on standard consumer hardware.

The geodesic lengths were predicted using a classifier, where each length represents one class. Initial experiments utilizing continuous regression heads failed to estimate geodesic lengths accurately. This failure is hypothesized to stem from the discrete, non-smooth geometry of algebraic word spaces, where smooth continuous loss functions struggle to model the abrupt length transitions caused by minor character edits. Reframing the task as discrete classification successfully resolved this optimization instability.

Nevertheless, architectural limitations remain at sequence length extremes. Geodesic length predictions for very short words exhibit higher error rates, an artifact of the sequence chunking mechanism producing graphs with too few nodes to establish rich topological representations. However, this error margin is practically negligible for the targeted cryptographic applications. Conversely, evaluating significantly longer words is constrained by the heavy computational bottlenecks of training data generation. To alleviate this constraint, future iterations could implement supergraphs that bundle multiple nodes to compress the spatial representation.

To further enhance classification accuracy across broad length regimes, a hierarchical multi-stage classification framework offers a promising path forward. In this setup, a coarse primary model first routes an input word into a specific length bracket, after which a specialized secondary model optimized strictly for that length range determines the exact geodesic length.

The computational workloads for this study were predominantly executed on a single Quadro 8000 GPU, with some supplementary data generation outsourced to workstations equipped with GeForce RTX 5080 cards. Expanding access to advanced AI clusters and distributed computing resources would drastically accelerate the dataset generation pipeline and allow for deeper scaling.

Ultimately, these findings highlight a highly promising direction for future work integrating machine learning with geometric group theory. The adaptability of WPNet suggests that similar neural constructions could be successfully applied to a wide range of algebraic challenges, including the conjugacy problem or for heuristic predictions to the shortest conjugator as recently discussed in (\cite{bridson2026conjugatorlengthfinitelypresented}). Furthermore, the structural embeddings of algebraic words into the continuous hypersphere generated by these models present a mathematically rich latent space that warrants independent geometric study.

\section{Conclusion}

This research establishes a highly promising direction for analyzing infinite groups using machine learning, presenting WPNet as an effective heuristic solver for the word problem and geodesic length prediction. By transforming unreduced algebraic words into dynamic graph structures, the architecture successfully clusters equivalent elements in a continuous embedding space without executing discrete reduction steps. While exact mathematical solvers suffer from exponential memory blowup on complex topologies, the presented approach achieves rapid evaluation times. Although predictions are heuristic and currently limited to word lengths up to 20,000, this scope comfortably encompasses the domains practically utilized in post-quantum cryptography. The successful deployment against the Wagner-Magyarik cryptosystem practically demonstrates the cryptographic severity of these structural embeddings. Future work can seamlessly adapt this flexible framework to a broader range of infinite groups and related algebraic challenges, such as the conjugacy problem. Furthermore, implementing multi-stage classifiers and supergraphs to bundle nodes will efficiently resolve existing computational bottlenecks at extreme sequence lengths.

\section{Acknowledgments}

The author would like to thank Delaram Kahrobaei and Ramon Flores for inspiring and fruitful discussion. Further, the author would like to thank the technical staff at Middlesex for outstanding support with access to the server infrastructure that made this work possible.

\bibliographystyle{plain} % Options: plain, abbrv, unsrt, alpha, etc.
\bibliography{references}

\end{document}